\documentclass[aps,prd,groupedaddress,amsmath,amssymb,amsfonts,nofootinbib,preprint,hidelinks]{revtex4-1}
\usepackage{amssymb}
\usepackage{amsmath, amssymb, verbatim, url, mathrsfs}
\usepackage{mathrsfs}
\usepackage{dutchcal}
\usepackage{mathtools}
\usepackage{verbatim}
\usepackage{amsthm}
\usepackage{framed}
\usepackage{wasysym}
\usepackage{upgreek}
\usepackage{color}
\usepackage[dvipsnames]{xcolor}
\usepackage{tensor}
\usepackage{accents}
\usepackage{dsfont}
\usepackage[colorlinks,linkcolor=blue,citecolor=blue]{hyperref}
\usepackage{enumerate}
\usepackage[normalem]{ulem}
\usepackage{longtable}
\usepackage{mathtools}
\usepackage{bm}
\usepackage{float}  

\definecolor{grey}{rgb}{0.4,0.4,0.4}
\definecolor{dullmagenta}{rgb}{0.4,0,0.4}
\definecolor{darkblue}{rgb}{0,0,0.4}
\definecolor{midblue}{rgb}{0,0,0.5}
\definecolor{midred}{rgb}{0.5,0,0}
\definecolor{orange}{rgb}{1,0.5,0}
\definecolor{lightbrown}{rgb}{0.75,0.5,0.25}
\definecolor{tan}{cmyk}{0.14,0.42,0.56,0}
\definecolor{djunglegreen}{cmyk}{0.99,0,0.52,0}
\definecolor{lightgreen}{rgb}{0,1,0}
\definecolor{olivegreen}{cmyk}{0.64,0,0.95,0.40}
\definecolor{midgreen}{rgb}{0.0,0.675,0.0}
\definecolor{darkgreen}{rgb}{0,0.5,0}

\numberwithin{equation}{section}

\newtheorem{theorem}{Theorem}[section]

\theoremstyle{definition}

\newcommand{\ca}{\mathcal{a}}
\newcommand{\ba}{\bar{\mathcal{a}}}
\newcommand{\cb}{\mathcal{b}}

\newcommand{\cc}{\mathcal{c}}

\newcommand{\bc}{\bar{\mathcal{c}}}

\newcommand{\mf}{\beta}
\newcommand{\mg}{\gamma}

\newcommand{\vertiiii}[1]{{\left\vert\kern-0.25ex\left\vert\kern-0.25ex\left\vert\kern-0.25ex\left\vert #1 \right\vert\kern-0.25ex\right\vert\kern-0.25ex\right\vert\kern-0.25ex\right\vert}}
\newcommand{\vertiii}[1]{{\left\vert\kern-0.25ex\left\vert\kern-0.25ex\left\vert #1 \right\vert\kern-0.25ex\right\vert\kern-0.25ex\right\vert}}

\newcommand{\tv}{\tilde{v}}

\newcommand{\trho}{\tilde{\rho}}

\newcommand{\tphi}{\tilde{\phi}}

\newcommand{\rphi}{\mathring{\phi}}
\newcommand{\rp}{\mathring{p}}
\newcommand{\rrho}{\mathring{\rho}}
\newcommand{\rv}{\mathring{v}}

\newcommand{\Rbb}{\mathbb{R}}

\newcommand{\del}[1]{{\partial_{#1}}}

\newcommand{\AND}{{\quad\text{and}\quad}}

\newcommand{\be}{\begin{equation}}
	\newcommand{\ee}{\end{equation}}

\allowdisplaybreaks

\DeclareMathOperator{\sgn}{sgn}

\begin{document}

\title[Fully nonlinear gravitational instabilities]{Fully nonlinear gravitational instabilities for expanding Newtonian universes with inhomogeneous pressure and entropy: Beyond the Tolman's solution}

\author{Chao Liu}
\email{chao.liu.math@foxmail.com}
\affiliation{Center for Mathematical Sciences and School of Mathematics and Statistics, Huazhong University of Science and Technology, Wuhan 430074, Hubei Province, China.}

%\author{Chao Liu}
%\email{chao.liu.math@foxmail.com}
%\affiliation{Center for Mathematical Sciences and School of Mathematics and Statistics, Huazhong University of Science and Technology, Wuhan 430074, Hubei Province, China.}

%\date{\formatdate{\day}{\month}{\year}, \currenttime}

\begin{abstract}  
Nonlinear gravitational instability is a crucial way to comprehend the clustering of matter and the formation of nonlinear structures in both the Universe and stellar systems.  However, with the exception of a few exact particular solutions for pressureless matter, there are only some  approximations and numerical and phenomenological approaches to study the nonlinear gravitational instability instead of mathematically rigorous analysis.
We construct a family of particular solutions of the Euler--Poisson system that exhibits the \textit{nonlinear} gravitational instability of \textit{matter with inhomogeneous pressure and entropy} (i.e., the cold center and hot rim) in the expanding Newtonian universe. Despite the density perturbations being homogeneous, the pressure is not, resulting in significant nonlinear effects.  
By making use of our prior work on   nonlinear analysis of a class of differential equations \cite{Liu2022b}, we estimate that the growth rate of the density contrast is approximately $\sim \exp(t^{\frac{2}{3}})$, much faster than the growth rate anticipated by classical linear Jeans instability ($\sim t^{\frac{2}{3}}$).
Our main motivation for constructing this family of solutions is to provide a family of reference solutions for conducting a fully nonlinear analysis of \textit{inhomogeneous} perturbations of density contrast.
We will present the general results in a mathematical article \cite{Liu2023b} separately.    Additionally, we  emphasize that our model does not feature any shell-crossing singularities before mass accretion singularities since we are specifically interested in analyzing the mathematical mechanics of a pure mass accretion model, which  poses limitations on the applicability of our model for understanding the realistic nonlinear structure formation.

\end{abstract}	

 
\maketitle
	%	\tableofcontents
	
%	\setcounter{tocdepth}{2}
	
%	\pagenumbering{roman} \pagenumbering{arabic}

\section{Introduction}
Gravitational instability characterizes the mass accretions of self-gravitating systems, clustering of matter and helps us understand the formations of stellar systems and the nonlinear structures in the universe. It traces back to Jeans \cite{Jeans1902} for Newtonian gravity in $1902$ (thus called \textit{Jeans instability}). 
However, it is worth noting Jeans' work is only in the \textit{linear regime} since he linearized the Euler--Poisson system.  It was generalized to general relativity by Lifshitz \cite{Lifshitz1946} and extended to the expanding universe by Bonnor \cite{Bonnor1957}, and later the linearized Jeans instability is widely applied (see \cite{Zeldovich1971,ViatcehslavMukhanov2013}). 
However, the linear Jeans instability has some inconveniences. The \textit{first inconvenience} comes from the linearization of the Euler--Poisson system. Due to the linearizations, the linear Jeans instability can be only applied to the case with \textit{small perturbations} of the uniform density distribution (i.e., the density contrast $\varrho:=(\rho-\rrho)/\rrho<1$) and only for a  time before the perturbations growing large, since the larger perturbations will lead to larger deviations from the linearized scheme. 
With the accretions of the mass, the derivations of the \textit{linear} Jeans instability will be completely spoiled since the increasing density leads to significant derivations from the linear regime. The \textit{second inconvenience} is the growth rate of the density contrast predicted by the classical linearized version of the Jeans instability can not yield the observed large inhomogeneities of the universe nowadays and formations of galaxies, because this growth rate ($\sim t^{\frac{2}{3}}$, see \cite{Bonnor1957,Zeldovich1971,ViatcehslavMukhanov2013,Liu2022}) is too slow and thus is much less efficient (see also \cite{Zeldovich1971,ViatcehslavMukhanov2013}). Therefore, it is urgent to \textit{study the fully nonlinear Jeans instability} and, as pointed out by Rendall \cite{Rendall2002} in 2002,  there are \textit{no results} on Jeans instability available for the \textit{fully nonlinear case}, and it becomes a long-standing open problem. The \textit{main goal} of this paper is to construct a family of gravitationally unstable solutions with homogeneous density, and  inhomogeneous pressure and entropy distributions, which serves as a family of \textit{reference solutions} for the fully nonlinear analysis of solutions with \textit{slightly inhomogeneous density}. We will present this result separately for \textit{slightly inhomogeneous density} in a mathematical article \cite{Liu2023b}.

On the other hand, although it is worthy to understand the fully nonlinear Jeans instability and the mathematical mechanics of a pure mass accretion model, it is important to acknowledge that the model presented in this article is idealized and simplified, lacking shell crossings and other singularities. In reality, things are not as simple as this. Usually, the evolution of density perturbations is believed to occur primarily in the linear regime because initial perturbations are small and require considerable time to grow. Then once the perturbation becomes of order unity, comparing with our model, the most cases are that the nonlinear approach may be of interest for only a short period due to shell-crossing causing violent relaxation and breaking down fluid approximations.  In such cases, N-body simulations have been applied in modern cosmology.

In fact, the model presented in this article can approximate some local portion of a giant gas cloud. Its density is almost homogeneous in this portion and the center of it is very \textit{cold} but the rim of the cloud is extremely \textit{hot}. In addition,  all the fluxes by the thermodynamics forces are negligibly small.  We further \textit{idealize} this model by assuming
the cloud initially has homogeneous density; the initial temperature distribution of this cloud is spherical symmetric and proportional to the square of the radius of the position (the accurate descriptions in \S\ref{s:model}). 
The \textit{key result} of this article is that, by using the mathematical tools developed in our previous paper \cite{Liu2022b} and taking the \textit{full nonlinear effects} into account, the growth rates of the density contrast are at least of order $\sim \exp(t^{\frac{2}{3}})$ (see later \eqref{e:mest1} for detailed expressions) or even blow up at the finite time (see \eqref{e:mest2}). This is much faster than that given by the classical linearized Jeans instability ($\sim t^\frac{2}{3}$, see \cite[\S $6.3$]{ViatcehslavMukhanov2013}). It may contribute to the explanations of the observed large inhomogeneities of the universe nowadays and the formations of galaxies.  As Peebles \cite[Chap. $1$, \S$4.$B]{Peebles2020} pointed out the exponential growth rates of the density contrast ``often has been cited as what is wanted.''  Although he also claim the exponential growth rates is not possible for the Einstein-de Sitter model, in our current model, the exponential growth does happen.

Some nonlinear strategies involving approximations and numerical methods (e.g., the famous \textit{Zel'dovich solutions}) have been discussed in several references (see  \cite{Zeldovich1971,ViatcehslavMukhanov2013,Arbuzova2014,Sciama1955}).
Another famous \textit{exact} solution describing the evolutions and collapses of the inhomogeneity is the \textit{Tolman solution} (see \cite[\S$6.4.1$]{ViatcehslavMukhanov2013} and \cite{LANDAU1975}) which gives an exact spherical symmetric \textit{dust} solution, but it can not be generalized to include the nonvanishing pressure effects, and the inconvenience of the parametric form of the solution is too complex to visualize the actual behaviors of it. 

In addition, \textit{the most important} thing, comparing with the Tolman solution, is that our method is robust. It can allow the presence of the pressure and can be generalized to more general cases by studying the corresponding dominant equations (e.g. the ordinary differential equation (ODE)  \eqref{e:feq0}) of the reference solutions first, then near every reference solutions, we can perturb the density contrast to obtain more general solution with inhomogeneous density as we will present in \cite{Liu2023b}. This paper is an example stating this idea and we will present this idea with other unstable models in future.   
Additionally, we  emphasize that our model does not feature any \textit{shell-crossing singularities} (see \S\ref{s:condis} for the proof) before mass accretion singularities since we are specifically interested in analyzing the mathematical mechanics of a pure mass accretion model.  
Therefore, it is important to recognize its limitations on the applicability of our model for understanding realistic nonlinear structure formation in the universe, since most of the outcomes of nonlinear evolutions, as demonstrated by \textit{N-body simulations} in modern cosmology, is the breakdown of the fluid approximation due to shell-crossing, which goes beyond the scope of the calculations presented in this paper.

\section{Models and assumptions}\label{s:model}
We use the Newtonian universe as an approximation of a local universe. Under the following assumptions, although the universe has the homogeneous density,
%\footnote{Our mathematical article \cite{Liu2023b} in preparation is for slightly inhomogeneous density too. }, 
it is inhomogeneous for the distributions of the pressure and entropy. It is of course an ideal model due to these perfect assumptions, but we want to develop a method for fully nonlinear gravitational instability \textit{with effective pressure resisting the gravity}. 
We will prove, in either case, the growth rate (at least $\sim \exp(t^\frac{2}{3})$, see \eqref{e:mest1}) of the density contrast due to the nonlinear Jeans instability is way faster than the one predicted by the classical linear Jeans instability ($\sim t^{\frac{2}{3}}$).

For the nonlinear version of the Jeans instability, we can not use Fourier analysis to solve the nonlinear differential equations derived from the Euler--Poisson system (see \cite{Liu2022} for alternative non-Fourier based proof of the linear Jeans instability). 
In \cite{Liu2022b}, we developed some preparing techniques for a class of the nonlinear ODEs and hyperbolic equations for the nonlinear Jeans instability. We intend to apply it to conclude the result of this article.

Let us now give the assumptions (and remarks for detailed meanings) of the model. According to the \textit{non-equilibrium thermodynamics} (see, for example, \cite[eqs. II.$(5)$, II.$(19)$ and   III.$(19)$]{DeGroot2012} or \cite[\S$69$]{Wang2005} ),  we assume the followings: 
\begin{enumerate} 
	\item[A1:] The fluids filled in the Newtonian universe are the ideal fluids and there is no chemical reactions, thus all the \textit{viscosity coefficients }and \textit{chemical affinities} of reactions vanish; All the \textit{phenomenological coefficients} $L_{ij}$ (see  \cite[Chap.IV]{DeGroot2012} for the definitions) are relatively small and negligible during the considered process, i.e., we assume all the $L_{ij}=0$, then all the fluxes in the entropy production vanish simultaneously with the thermodynamic forces (due to the facts that the entropy production satisfies $\sigma=\sum_k J_k\cdot X_k$ and the linear phenomenological law gives $J_i=\sum_kL_{ik}X_k$).   Therefore, every comoving parcel is adiabatic (isentropic). 
	\item[A2:] The initial entropy is distributed proportionally to $|\boldsymbol{x}|^2$, while the adiabatic comoving parcel ensures that the entropy remains proportional to the comoving position $|\boldsymbol{q}|^2$.
\end{enumerate}

In other words, these assumptions means the Newtonian universe can be described by the following \textit{reduced Euler--Poisson system} (we use
the Einstein summation convention), 
\begin{gather}
	\del{t}\rho+\del{i}(\rho v^i) = 0 ,\label{EP1}\\
	\del{t}v^i +v^j\del{j} v^i+\frac{\del{}^i p}{\rho}+\del{}^i\phi = 0 ,\label{EP2}\\
	\del{t}S+v^i\del{i}S=0, \label{EP3a} \\
	\Delta\phi=\delta^{ij}\del{i}\del{j} \phi =  4\pi G\rho,    \label{EP3}
\end{gather}
where  $\rho$, $ v^i$, $p$, $\phi$ and $S$ are the density, velocities, pressure of the fluids, gravitational potential and specific entropy, respectively. 
The \textit{equation of state} is presumed by 
\begin{equation}\label{e:eos1}
	p = K e^{\frac{S-S_0}{c_V}}  \rho^{\gamma} +\mathfrak{p} , \quad \text{for} \;  \gamma=\frac{4}{3} \AND K\geq 0 .  
\end{equation}
where $\mathfrak{p}\in \Rbb$ is a constant. 
The initial data at $t=t_0$ are given by
\begin{gather}
	\rrho(t_0)= \frac{\iota^3}{6\pi Gt_0^2},  \quad  \rv^i(t_0,x^k)=\frac{2}{3t_0}x^i, \label{e:dataa1} \\ \rphi(t_0,x^k)=\frac{2}{3}\pi G\rrho(t_0) \delta_{ij} x^i x^j \AND S(t_0, x^k)= S_0+ c_V \ln (\kappa t_0^{-\frac{4}{3}} \delta_{kl} x^k x^l)^{\sgn(1-\iota^3)} , \label{e:datab1}
\end{gather}
where $\kappa>0$ is a constant and $\sgn$ is a \textit{sign function}\footnote{The sign function implies $\sgn(\alpha)=1$ if $\alpha>0$, and $\sgn(\alpha)=0$ if $\alpha=0$. }, 
and $\iota$ is a constant determined by   
\begin{equation*}\label{e:iota}
	\iota:=	\iota(\tilde{K})= 
	\Bigl( \frac{1}{2}\sqrt{1+	18\tilde{K}} +\frac{1}{2}\Bigr)^{\frac{1}{3}} - \Bigl(\frac{1}{2}\sqrt{1+18	\tilde{K}}-\frac{1}{2}\Bigr)^{\frac {1}{3}}\in (0,1]  \AND \tilde{K}:=\frac{K^3\kappa^3}{\pi G } ,
\end{equation*}

Before proceeding, about this model, let remark some notable facts: 
\begin{enumerate}
	\item Note $\tilde{K}$ and $\iota$ are \textit{dimensionless constants} depending on the \textit{molar mass} of the fluids and the distributions of the entropy or temperature (see Appendix \ref{s:nondim} for details). 
	\item If $\tilde{K}=0$ (equivalently, $\iota=1$), then this model reduces to an isentropic case $S(t_0, x^k)\equiv S_0$ with a constant pressure $\mathfrak{p}$, thus the data and the solution \eqref{e:exsol1}--\eqref{e:exsol2} given below reduce to the classical Newtonian solutions for the homogeneous and isotropic Newtonian universe given in, e.g.,  \cite[\S$1.2.3$]{ViatcehslavMukhanov2013} or \cite[\S$10.2$]{Zeldovich1971}. The results of this article reduce to the case of Tolman solutions. 
	\item If $\iota \neq 1$, then the data of the entropy \eqref{e:datab1} implies the initial distribution of the temperature $\mathcal{T} \propto |\boldsymbol{x}|^2$ (see Appendix \ref{s:nondim} for detailed explanations).  
	\item In the equation of state \eqref{e:eos1}, the inclusion of the term $\mathfrak{p}$ does not alter the mathematical derivations; rather, it is included to increase the generality of the model. 
	\item	Note that $\iota$ satisfies an important identity (crucial in later derivations), 
	\begin{align}\label{e:ioeq}
		\iota^3+ 9 \Bigl(\frac{\tilde{K}}{6}\Bigr)^{\frac{1}{3}}   \iota -1=0 ,  
	\end{align}
	and $\iota(\tilde{K})$ is a decreasing function\footnote{Since one can verify that its derivative
		\begin{align*}
			\iota^{\prime}(\tilde{K})=\frac{3 }{\sqrt[3]{2} \sqrt{18 \tilde{K}+1}}\biggl(\frac{1}{\bigl(\sqrt{18 \tilde{K}+1}+1\bigr)^{2/3}}-\frac{1}{\bigl(\sqrt{18 \tilde{K}+1}-1\bigr)^{2/3}}\biggr) <0. 
	\end{align*}}, and $
		\lim_{\tilde{K}\rightarrow 0} \iota(\tilde{K})=1$, and $ \lim_{\tilde{K}\rightarrow +\infty} \iota(\tilde{K})=0$. 
%	\item
%	The initial data of pressure, by \eqref{e:eos1}--\eqref{e:datab1}, is given by $ \ring{p}(1) 		=K  \kappa  \delta_{kl} x^k x^l \rrho(1)^{ \frac{4}{3}}+\mathfrak{p}$. 
\end{enumerate}

To simplify calculations, letting $s=(S-S_0)/c_V$ and along with the non-dimensionalizations in Appendix \ref{s:nondim}, we proceed with  the \textit{dimensionless} and \textit{normalized} Euler--Poisson system: 
\begin{gather}
	\del{t}\rho+\del{i}(\rho v^i) = 0 ,\label{EP1}\\
	\del{t}v^i +v^j\del{j} v^i+\frac{\del{}^i p}{\rho}+\del{}^i\phi = 0 ,\label{EP2}\\
	\del{t}s+v^i\del{i}s=0, \label{EP3a} \\
	\Delta\phi=\delta^{ij}\del{i}\del{j} \phi =  4\pi \rho.     \label{EP3}
\end{gather}
The \textit{equation of state} becomes
\begin{equation}\label{e:eos}
	p=K e^{s} \rho^{\frac{4}{3}}+\mathfrak{p}, \quad \text{for} \;  K\geq 0 .  %\AND K\in \Bigl(0,\frac{1}{3}\Bigr)
\end{equation}  
The initial data at $t=1$ is given by
\begin{gather}
	\rho|_{t=1}= \frac{\iota^3}{6\pi},  \quad  v^i|_{t=1}=\frac{2}{3 }x^i \AND  s|_{t=1}=  \ln (  \delta_{kl} x^k x^l)^{\sgn(1-\iota^3) } .   \label{e:datab}
\end{gather}  

Let us try to find a homogeneous and expanding Newtonian solution in the meaning of a homogeneous density and the \textit{Hubble law} dominated velocity field, we then obtain
\begin{equation}
	\rho(t,x^k)=\rrho(t),\quad
	v^i(t,x^k)=\rv^i(t,x^k)=H(t)   x^i.\label{eq 4}
\end{equation}
There is an exact solution to the Euler--Poisson system \eqref{EP1}--\eqref{EP3} and the data \eqref{e:datab} on $(t,x^k) \in [t_0,\infty)\times \Rbb^3$, 
\begin{gather}
	\rrho(t)= \frac{\iota^3}{6\pi  t^2}, \quad \rp(t) = K t^{-\frac{4}{3}} \delta_{kl} x^k x^l  \rrho^{\frac{4}{3}}+\mathfrak{p},  \quad
	 \rv^i(t,x^k)=\frac{2}{3t}x^i, \label{e:exsol1} \\ \rphi(t,x^k)=\frac{2}{3}\pi  \rrho \delta_{ij} x^i x^j =\frac{\iota^3}{9t^2}  \delta_{ij} x^i x^j  \AND \mathring{s}(t,x^k)=\ln (t^{-\frac{4}{3}} \delta_{kl} x^k x^l)^{\sgn(1-\iota^3) },    \label{e:exsol2}
\end{gather}

\section{Main results and ideas}\label{s:stp} 
This article intends to conclude the nonlinear behavior of the homogeneous perturbations of the density contrast $\varrho:=(\rho-\rrho)/\rrho$ by the following two steps and the main results  are given by the following estimates \eqref{e:mest1}--\eqref{e:mest2} of the lower bounds of the growth rate of the density contrast.

\underline{Step $1$:} Let us  assume that $\mf$ and $\mg$ are two given \textit{positive} constants and that the initial data (at $t=1$) of \eqref{EP1}--\eqref{EP3} have an homogeneous initial perturbations and are  characterized by two  positive parameters $\mf$ and $\mg$ in the following ways, 
 \begin{equation} 	\rho|_{t=1}=(1+\mf)\frac{\iota^3}{6\pi} ,\quad 
 	 v^i |_{t=1}=\Bigl(\frac{2}{3 }-\mg \Bigr)x^i     \AND s|_{t=1}=\ln \bigl(  (1+\beta)^{\frac{2}{3}} \delta_{kl} x^k x^l\bigr)^{\sgn(1-\iota^3) } . \label{e:data2} 
 \end{equation}
Then we will prove the solution of the Euler--Poisson system \eqref{EP1}--\eqref{EP3}  becomes (we use notation $(\cdot)^\prime:=d(\cdot)/dt$)
	\begin{gather} 
	 \rho(t) =(1+f(t) ) \rrho(t)= \frac{\iota^3(1+f(t) )}{6\pi  t^2} ,  \label{e:sl1}\\
v^i  (t,x^i) =\frac{2}{3t}x^i - \frac{ f^\prime(t)}{3 (1+f(t))} x^i , \label{e:slv} \\
\phi(t,x^i) = \frac{2 }{3}\pi   \rrho (1+f(t)) |\boldsymbol{x}|^2= \frac{\iota^3(1+f(t)) |\boldsymbol{x}|^2}{9  t^{2}}   ,  \\
s(t, x^k)=\ln \bigl( t^{-\frac{4}{3}} (1+f)^{\frac{2}{3}} \delta_{kl} x^k x^l\bigr)^{\sgn(1-\iota^3) } .  \label{e:sl2}
\end{gather}
and the \textit{density contrast} $\varrho(t)=f(t)$ 
 	where $|\boldsymbol{x}|^2:=\delta_{ij} x^i x^j$ and $f(t)$   is a solution of the following nonlinear ODE,  
 	\begin{gather}
 	f^{\prime\prime}(t)+\frac{4}{3t} f^\prime(t)-\frac{2}{3 t^2} f(t)(1+f(t)) -\frac{4(  f^\prime(t))^2}{3(1+f(t))}= 0 ,\label{e:feq1b} \\
 	f|_{t=t_0}=\mf \AND f^\prime|_{t=t_0}=3(1+\mf)\mg. \label{e:feq2b}
 	\end{gather}	
Moreover, the pressure becomes $p(t)=\frac{K  \iota^4  }{(6\pi )^{\frac{4}{3}}t^{4}}  (1+f)^{2} \delta_{kl} x^k x^l $.

\underline{Step $2$:} In Step $1$,  we have represented the perturbation solution in terms of functions $f(t)$ and its derivative $f_0(t):=f^\prime(t)$. To understand the behaviors of the perturbation solution, especially the growth rates of the density contrast $\varrho$, we have to know the detailed behaviors of the functions $f$ and $f_0$.  In fact, the behaviors of $f$ and $f_0$ can be acquired by solving the ODE \eqref{e:feq1b}--\eqref{e:feq2b} which has been well studied in our companion article \cite{Liu2022b}. We list the conclusions of the solutions to  the ODE \eqref{e:feq1b}--\eqref{e:feq2b} in Appendix \ref{s:ODE} and using it, we conclude the density contrast has the lower bound estimate, for $t\in(1,t_m) $, 
\begin{align}\label{e:mest1}
	&\varrho(t)=f(t)  >\notag  \\ 
&	\exp \biggl( \frac{3\bigl(\ln(1+\mf)+3\mg \bigr)  t^{\frac{2}{3} }+2\bigl(\ln(1+\mf)-\frac{9}{2} \mg \bigr)  t^{-1}}{5}\biggr)  - 1    . 
\end{align}
In addition, by Theorem \ref{t:mainthm0}, if further the initial data satisfies  
$\mg>  1/3$, we have an improved lower bound estimate on the growth rate of $\varrho$, for $t\in(1,t_m)$, 
\begin{align}\label{e:mest2}
	\varrho(t)=	f(t)>&\frac{1+\mf }{ \bigl(1-3\mg +  3\mg t^{-\frac{1}{3}} \bigr)^{3} } -1 .
\end{align}
The lower bound of $\varrho$ blows up at $t^\star=\bigl(1-\frac{1}{3\mg}\bigr)^{-3}>t_0=1$. These lower bounds give an estimate of the growth rates of the density contrast $\varrho$.

In the rest of this article, we will only need to elaborate Step $1$, i.e., solving the Euler--Poisson system \eqref{EP1}--\eqref{EP3} under the perturbed data \eqref{e:data2} and further the perturbation equations.

%-----------------------Sec 2-----------------------

\section{Equations of perturbations}  
Let us first decompose the variables $(\rho,v^i,p,\phi)$ to the exact background solution \eqref{e:exsol1}--\eqref{e:exsol2} and the perturbed parts, and define a density contrast $\varrho$, 
\begin{gather}\label{e:perv}
	\rho =\rrho + \trho,\quad
	v^i= \rv^i +\tv^i,\quad \phi=\rphi + \tphi, \quad s=\mathring{s}+\tilde{s} \\
	p =\rp +  \tilde{p} , \AND \varrho:= \frac{\trho}{\rrho}  .  \label{eq 6}
\end{gather}

Next we introduce the Lagrangian coordinates $q^k$ defined by $ x^k=a(t)q^k $ where\footnote{In fact, $a(t)=a(1) t^{\frac{2}{3}}=  t^{\frac{2}{3}}$ provided $a(1)=1$, since by the Hubble law \eqref{eq 4} and the Lagrangian coordinates $x^k=a(t)q^k$, we obtain  $H(t):=\frac{\dot{a}(t)}{a(t)}$. Then by $H=\frac{2}{3t}$ (see \eqref{eq 4} and \eqref{e:exsol1}), we can solve $a(t)$. } $a(1):=1$, and the time derivatives are obtained at $q^k$ (i.e., the material derivatives).
We also denote
\begin{gather} 
	D_t:= \del{t}|_{q^k}=\del{t}|_{x^k}+\rv^ i \del{i}=\del{t}|_{x^k}+Hx^j   \del{j}  , \label{e:lag2} \\
	D_i:=a(t)\del{i} . \label{e:lag3}
\end{gather} 
Note in the next, we will slightly abuse the notations and do not distinguish the variables in terms of the Eulerian $x^i$ and Lagrangian coordinate $q^i$, that is, we abuse, e.g.,  $\tv^k(t,x^i)$ and $\tv^k(t,q^i)$ for the simplicity of the expressions and readers should be clear according to the contexts. 

Let us review how to reexpress the Euler--Poisson system \eqref{EP1}--\eqref{EP3} in terms of the perturbation variables $(\varrho, \tv^i, \tilde{s},\tphi)$ given by \eqref{e:perv}--\eqref{eq 6}.  
First, let us consider the conservation of mass. Substituting the decomposition \eqref{e:perv}--\eqref{eq 6} into Eq.  \eqref{EP1},  
using the Hubble laws \eqref{eq 4} 
and
applying the Lagrangian coordinates \eqref{e:lag2}--\eqref{e:lag3}, the conservation of mass   
\eqref{EP1} becomes 
\begin{equation*}%\label{e:mass2}
	D_t \trho  + 3H\trho  +\rrho a^{-1} D_i  \tv^i +\trho a^{-1} D_i \tv^i  +\tv^i  a^{-1} D_i  \trho =  0 .
\end{equation*} 
Using \eqref{eq 6} and $\del{t} \rrho+3H\rrho=0$, and after straightforward calculations, we obtain 
\begin{equation}\label{e:mass3}
	D_t \varrho   +  (1+ \varrho) a^{-1} D_i \tv^i  + \tv^i  a^{-1} D_i  \varrho =  0  . 
\end{equation}

Second, by \eqref{eq 6} and \eqref{e:lag3},  the Poisson equation in terms of the Lagrangian coordinate $q^i$ becomes  
\begin{equation}\label{e:ps2}
	\delta^{ij}D_iD_j \tphi %=   4\pi a^2 G\trho 
	=   4\pi a^2   \rrho \varrho =   \frac{2 \iota^3}{3t^{\frac{2}{3}}} \varrho .  
\end{equation}

In the end, we turn to the balance of momentum \eqref{EP2}.  
By subtracting the background \eqref{e:exsol1}--\eqref{e:exsol2} from \eqref{EP2},  and
in terms of Lagrangian coordinates, using \eqref{e:lag2}--\eqref{e:lag3}, with the help of \eqref{eq 6}, the balance of momentum \eqref{EP2} becomes 
\begin{equation}\label{e:veq1}
	D_t  \tv^i   +  H \tv^i+   \tv^j  a^{-1} D_j \tv^i+\frac{a^{-1} D^i  (\tilde{p} / \rrho)}{1 +\varrho} - \frac{2 K  \iota}{(6\pi )^{\frac{1}{3}}t^{\frac{4}{3}}} \frac{   \varrho}{1+\varrho}   q^i + a^{-1} D^i  \tphi =  0 .
\end{equation}

Then we consider the conservation of the entropy. In terms of Lagarangian coordinates, direct calculations that imply \eqref{EP3a} become   
\begin{align} \label{e:seq2}
	D_t \tilde{s}+ \sgn(1-\iota^3) \cdot \frac{2  \delta_{ij} q^i \tilde{v}^j}{a \delta_{kl} q^k q^l } +\tv^i a^{-1} D_i \tilde{s} = 0 .
\end{align}

Gathering above equations \eqref{e:mass3}--\eqref{e:seq2} together, the Euler--Poisson system \eqref{EP1}--\eqref{EP3}, in terms of the Lagrangian coordinates, becomes
\begin{gather}
	D_t \varrho   +  \frac{(1+ \varrho)}{a}  D_i \tv^i  + \frac{\tv^i}{a}   D_i  \varrho =  0 , \label{e:EP.a.1} \\ 
	D_t  \tv^i   +  H \tv^i+   \tv^j  a^{-1} D_j \tv^i+\frac{a^{-1} D^i  (\tilde{p} / \rrho)}{1 +\varrho} - \frac{2   K  \iota}{(6\pi )^{\frac{1}{3}}t^{\frac{4}{3}}} \frac{   \varrho}{1+\varrho}   q^i + a^{-1} D^i  \tphi =  0 , \label{e:EP.a.2} \\
	D_t \tilde{s}+\sgn(1-\iota^3) \cdot\frac{2\delta_{ij} q^i \tilde{v}^j}{a \delta_{kl} q^k q^l } +\tv^i a^{-1} D_i \tilde{s} = 0 \label{e:EP.a.2a} \\
	\delta^{ij}D_iD_j \tphi   =  \frac{2 \iota^3}{3t^{\frac{2}{3}}} \varrho .  \label{e:EP.a.3}
\end{gather}

\section{Perturbation solutions}
In this section, let us focus on solving Eqs. \eqref{e:EP.a.1}--\eqref{e:EP.a.3} under the initial perturbation \eqref{e:data2}. In this assumption, we have that the density contrast is independent of the spatial variables and  the velocity satisfies the Hubble's law, thus we assume the forms of $\varrho$ and $\tv^i$ by
\begin{align}\label{e:vel}
	\varrho(t,q^k)\equiv \varrho(t) \AND \tv^i(t,q^k)=\tilde{H}(t)q^i 
\end{align}
where the function $\tilde{H}(t)$ is to be determined variable. 
In this case, by \eqref{e:EP.a.3}, direct calculations imply
\begin{equation}\label{e:phi}
	\tphi=\frac{2}{3}\pi a^2 \rrho \varrho|\boldsymbol{q}|^2=\frac{ \iota^3 \varrho |\boldsymbol{q}|^2}{9t^{\frac{2}{3}}} \AND \delta^{ij} D_i\tphi= \frac{4}{3} \pi a^2  \rrho \varrho q^j=\frac{2 \iota^3 \varrho }{9t^{\frac{2}{3}}} q^j  
\end{equation} 
where $|\boldsymbol{q}|^2:=\delta_{ij} q^i q^j$.

Now our task becomes solving $\varrho(t)$ and $\tilde{H}(t)$ from Eqs.  \eqref{e:EP.a.1} and \eqref{e:EP.a.2}. 
Taking the divergence of $\tv^i$ leads to $D_i\tv^i=3\tilde{H}(t)$, noting  \eqref{e:vel} yields $D_i\varrho=0$, \eqref{e:EP.a.1} implies 
\begin{equation}\label{e:vel3}
\tilde{H}(t)=-\frac{t^{\frac{2}{3}}}{3}  \bigl(\ln (1+\varrho(t))  \bigr)^\prime \quad \Rightarrow  \quad 	\tv^i =-\frac{t^{\frac{2}{3}}}{3}q^i \bigl(\ln (1+\varrho(t))  \bigr)^\prime =- \frac{t^{\frac{2}{3}} \varrho^\prime(t)}{3(1+\varrho(t))} q^i . 
\end{equation} 

By the data \eqref{e:datab} and \eqref{e:data2}, we have the data $\tilde{s}|_{t=1}= \ln(1+\beta)^{\frac{2}{3}\sgn(1-\iota^3)}$ which is a homogeneous perturbation of the entropy, thus we assume $\tilde{s}$ is also independent of $x^i$ and is homogeneous, then $D_i \tilde{s}=0$. Using the velocity \eqref{e:vel3}, direct integrating \eqref{e:EP.a.2a} imply  
 \begin{equation*}
 	\tilde{s}(t) =\ln [  (1+\varrho(t))^{\frac{2}{3}\sgn(1-\iota^3)}  ]  . 
 \end{equation*}

Next, we solve $\varrho(t)$. Noting, if $\iota\neq 1$, then $
	\tilde{p} =K \delta_{kl} q^k q^l\rrho^{\frac{4}{3} } [(1+\varrho)^2-1]=K \delta_{kl} q^k q^l\rrho^{\frac{4}{3} } \varrho (\varrho +2)
$,
and substituting \eqref{e:phi}--\eqref{e:vel3} into \eqref{e:EP.a.2}, straightforward calculations imply
\begin{equation*}
\tilde{H}^\prime +\frac{2}{3t} \tilde{H} + \tilde{H}^2 a^{-1} + \frac{2 K \iota \varrho  }{(6\pi  )^{\frac{1}{3}} t^{\frac{4}{3}}  }  +\frac{2\iota^3 \varrho}{9t^{\frac{4}{3}}}= 0. 
\end{equation*}
Then the \textit{crucial step}  is using the identity \eqref{e:ioeq}, i.e., $\iota^3+\frac{9K}{(6\pi)^{\frac{1}{3}}}\iota=1$ in terms of the dimensionless $K$, which leads to
\begin{equation}\label{e:Heq}
 \tilde{H}^\prime +\frac{2}{3t} \tilde{H} + \tilde{H}^2 a^{-1}   +\frac{2  \varrho }{9t^{\frac{4}{3}}}= 0. 
\end{equation}
If $\iota=1$, then $K=0$ and $p=0$. We also arrive at \eqref{e:Heq}. 

Noting that \eqref{e:vel3} yields
\begin{equation*}
\tilde{H}^\prime =-\frac{2}{9} \frac{t^{-\frac{1}{3} } \varrho^\prime}{1+\varrho} -\frac{t^{\frac{2}{3}}}{3} \frac{\varrho^{\prime\prime}}{1+\varrho} +\frac{t^{\frac{2}{3}}}{3} \frac{(\varrho^\prime)^2}{(1+\varrho)^2}, 
\end{equation*}
and substituting it and \eqref{e:vel3} into \eqref{e:Heq}, we arrive at\footnote{Note the same equation has also been obtained in \cite{Fosalba1998} for spherical collapse model. }
\begin{equation} \label{e:rhoeq}
\varrho^{\prime\prime}(t) +   \frac{4}{3t}  \varrho^\prime (t)   -      \frac{2}{3 t^2}  \varrho(t)  (1+ \varrho(t) )  -   \frac{4}{3}   \frac{( \varrho^\prime (t))^2 }{ (1+ \varrho(t))}     = 0 . 
\end{equation}

By the data \eqref{e:data2} and the definition of the perturbed variables \eqref{e:perv}--\eqref{eq 6}, we obtain the data of $\varrho|_{t=1}$, 
\begin{equation}\label{e:dtr}
	\varrho|_{t=1}=\mf . 
\end{equation}

Since \eqref{e:rhoeq} is a second order ODE, to solve this equation, we have to know the initial data of $\varrho^\prime|_{t=1}$. In order to find this data and solve the Euler--Poisson system \eqref{e:EP.a.1}--\eqref{e:EP.a.3}, we note that \eqref{e:EP.a.1} must hold at the initial time $t=1$. Thus, with the help of \eqref{e:data2} and the data $\tv^i|_{t=1}=(v^i-\rv^i)|_{t=1}= -\mg q^i$ (by \eqref{e:datab} and \eqref{e:data2}, and noting $a(1)=1$), noting $D_i\varrho=0$, we have
\begin{align}\label{e:dtdtr}
	\varrho^\prime|_{t=1}=-\Bigl(\frac{1+\varrho}{a}D_i \tv^i\Bigr)\Big|_{t=1}-\Bigl(\frac{\tv^i}{a}D_i\varrho\Bigr)\Big|_{t=1} =3(1+\mf)\mg. 
\end{align}

Using the ODE \eqref{e:rhoeq} and the data \eqref{e:dtr}--\eqref{e:dtdtr}, with the help of Theorem \ref{t:mainthm0} (see Appendix \ref{s:ODE}), we obtain
$\varrho(t)=f(t)$ where $f(t)$ is given by Theorem \ref{t:mainthm0}. 
Further, we conclude the solutions \eqref{e:sl1}--\eqref{e:sl2}. This completes Step $1$. In summary, we obtain a family of solutions depending on two parameters $\beta $ and $\gamma $.

\section{Conclusions and discussions}\label{s:condis}
Let us firstly compare the nonlinear growth rates \eqref{e:mest1}--\eqref{e:mest2} of the density contrast  with that ($\sim t^{\frac{2}{3}}$) predicted by the classical linear version of the Jeans instability. 
By using $t^{-1}\leq t^\frac{2}{3}$ for $\ln(1+\mf)-\frac{9}{2} \mg < 0$ and $t^{-1} >0$ for $\ln(1+\mf)-\frac{9}{2} \mg\geq 0$, \eqref{e:mest1} implies 
\begin{equation}\label{e:mest3}
	\varrho(t)  > \exp \bigl(   A t^{\frac{2}{3} }\bigr)  - 1   = At^{\frac{2}{3}}+ \mathrm{O}(t^{\frac{4}{3}})
\end{equation}
where $A:=\min\{\ln(1+\mf), \frac{3}{5}\bigl(\ln(1+\mf)+3\mg \bigr)\}$ is a constant and $\mathrm{O}(t^{\frac{4}{3}})$ means the remainder terms are at least of order $t^{\frac{4}{3}}$. 
We note the growth rate ($\sim t^{\frac{2}{3}}$) by the classical Jeans instability is just the first order approximation of the lower bound estimate \eqref{e:mest3} if expanding $\exp \bigl(   A t^{\frac{2}{3} }\bigr) $ with respect to $t^\frac{2}{3}$. The \textit{nonlinear effects indeed significantly boost the growth} of the density contrast as $\varrho$ grows larger.  
According to the Taylor expansion \eqref{e:mest3}, we see that, when $t$ is small enough, it consists of the result of the classical linearized Jeans instability. From these improved faster growth rates due to the nonlinear effects, we can see that indeed the classical linearized  Jeans instability can only be applied to the small initial perturbation of the density and only work for a short time before the density grows large enough. However, the nonlinear method proposed by this article does not require the initial perturbations small and it works for a long time before the Euler--Poisson system breaks down.

The method of this article relies on the nonlinear analysis of a type of nonlinear differential equations, which is mathematically rigorous without approximations and numerical calculations. This method is \textit{robust and systematic},  and if we also use the ideas and method (the Cauchy problem of the Fuchsian formulation of a second order hyperbolic equation which allows certain pressure) from our previous paper \cite[\S$3$]{Liu2022b}, it is possible to study the general cases of the nonlinear Jeans instability, at least for the case with nonvanishing pressure and small inhomogeneous perturbations.  

Although this model of the universe in the present paper is a simplified model, it can still capture some of the main nonlinear effects on the local Newtonian universe. Thus this result helps us have a better understanding of the \textit{formation} of the \textit{nonlinear structures} in the \textit{universe} and \textit{stellar systems}.  The lower bound estimates  \eqref{e:mest1}--\eqref{e:mest2}  of the growth rates of the density contrast are far more efficient than the one predicted by the classical linear Jeans instability. Therefore, it may be possible that these nonlinear results of the growth rates can contribute to the explanations of the observed large inhomogeneities of the universe nowadays and the formations of galaxies. Due to these much larger and more efficient growth rates of the density contrast, we may not require substantial initial perturbations of the density contrast at the early times of the universe and weaken the constraints on the initial spectrum of perturbations. 

In the end, we would like to note that there are no shell-crossing singularities\footnote{See, for instance,  \cite{Frauendiener1995,Szekeres1999} for related cases. } present in our model. This is evident when observing the velocity field \eqref{e:slv} in spherical coordinates, 
\begin{equation*}
	v  (t,r) =\Bigl(\frac{2}{3t}  - \frac{ f^\prime(t)}{3 (1+f(t))}\Bigr) r
\end{equation*}	
where $r:=|\boldsymbol{x}|$. 	
This velocity field implies 
	\begin{equation*}
		\del{r} v =   \frac{2}{3t}   - \frac{ f^\prime(t)}{3 (1+f(t))}   . 
 	\end{equation*}
Using relations in the companion article \cite[eqs. $(3.18)$ and $(3.89)$]{Liu2023b}, i.e,
\begin{equation*}
	\frac{(f^\prime)^2}{(1+f)^2} =\frac{\chi f}{Bt^2} \AND\frac{	\chi(t)}{B}= 4  + \frac{ \mathfrak{G}(t) }{B} \Rightarrow  \frac{f^\prime}{1+f} =\frac{2}{t} \sqrt{f} \Bigl(1+\frac{\mathfrak{G}}{4B}\Bigr)^{\frac{1}{2}}  , 
\end{equation*} 
we conclude
\begin{equation*}
	\lim_{t\rightarrow t_m} \del{r} v  =  \lim_{t\rightarrow t_m} \frac{2}{3t} \biggl[1-\sqrt{f} \Bigl(1+\frac{\mathfrak{G}}{4B}\Bigr)^{\frac{1}{2}} \biggr] =-\infty \AND |\del{r} v| <\infty \;\text{for}\; t\in[1,t_m),
\end{equation*}
since $\lim \sqrt{f}/t=\infty$ by Theorem \ref{t:mainthm0}.$(3)$ and $\lim_{t\rightarrow t_m}\mathfrak{G}(t)=0$ (by \cite[Proposition B.$4$]{Liu2023b}). This result implies the characteristic curves generated by the velocity field do not intersect for $t\in[1,t_m)$ until $t=t_m$ (the mass accretion singularities lie on the whole $\{t_m\} \times \Rbb^3$) and it means there is no shell crossing singularities for $t\in[1,t_m)$.

In summary, this model gives mathematically rigorous and physically decent nonlinear estimates on the growth rates (at least $\sim \exp(t^\frac{2}{3})$) of the density contrast $ \varrho$ on the local portion of the universe characterized by the Newtonian expanding universe.  
The mathematical tools and methods have the potential for  general cases of the nonlinear version of the Jeans instability. Based on these tools and methods developed by our prior article \cite{Liu2022b}, the fully nonlinear Jeans instabilities both for the Newtonian universe and general relativity are in progress.

%-------------------------------------------------------------New Section----------------------------------------------------------------------------------------

\appendix

\section{Non-dimensionalizations and parameters}\label{s:nondim} 
First note the dimensions of variables in this article are
\begin{gather*}
	[x^i]=L, \quad [t]=T \;\text{(Time)},\quad [\mathcal{T}]=\mathbb{T} \;\text{(Temperature)},\quad  [s]=1, \quad [p]=\frac{M}{LT^2}, \quad [\rho]=\frac{M}{L^3},  \\ [G]=\frac{L^3}{MT^2},\quad
	[\phi]=\frac{L^2}{T^2}, \quad [v^i]=\frac{L}{T}, \quad [K_T]=\frac{L^3}{T^2M^{\frac{1}{3}}} . 
\end{gather*}
To
introduce dimensionless variables, we define 
\begin{equation*} 
	\rho=\rho_T \hat{\rho}, \quad K= K_T \hat{K}  \AND \mathcal{T}=\mathcal{T}_T\hat{\mathcal{T}}
\end{equation*}
where  $\rho_T$, $K_T$ and $\mathcal{T}_T$ are the typical values for the density, the coefficient of the equation of state and the temperature. Then let
\begin{gather*}
	\hat{p}=\frac{p}{  K_T \rho_T^{\frac{4}{3}}}, \quad \hat{x}^i=\sqrt{\frac{G\rho_T}{K_T \rho_T^{\frac{1}{3}}}} x^i=\rho_T^{\frac{1}{3}} \sqrt{\frac{G }{K_T} } x^i, \quad \hat{t}=\sqrt{G\rho_T}t, \quad \hat{\phi}=\frac{1}{K_T \rho_T^{\frac{1}{3}}} \phi  , \\
	v^i= \sqrt{K_T\rho_T^{\frac{1}{3}}}\hat{v}^i  \AND   \kappa=\frac{G^{\frac{1}{3}}}{K_T}\hat{\kappa}  . 
\end{gather*}
Then $[\hat{K}] = [\hat{p}]=[\hat{v}^i]=[\hat{\rho}]=[\hat{x}^i]=[\hat{t}]=[\hat{\mathcal{T}}]=[\hat{\kappa}]=1$. 
Thus all our dynamical variables and coordinates are dimensionless and the three constants $\rho_T$, $K_T$ and $\mathcal{T}_T$ can
be used to fix the length, time and  temperature scales by using units so that
\begin{equation*}
	\rho_T=  \frac{1}{t_0^2 G},   \quad K_T= \frac{G^{\frac{1}{3}}}{\kappa} \AND \mathcal{T}_T=1
\end{equation*}
where $t_0$ is the initial time. In this case, we have
$	\hat{t}=t/t_0$ and $ \hat{\kappa}=1$. 

We \textit{claim} $\tilde{K}$ is dimensionless quantity and further so is $\iota$, since 
\begin{equation*}
	\tilde{K}=\frac{K^3\kappa^3}{\pi G }=\frac{K_T^3\hat{K}^3G\hat{\kappa}^3}{K_T^3\pi G }=\frac{ \hat{K}^3 \hat{\kappa}^3}{ \pi  } .
\end{equation*}

Recalling \cite[Chap. II]{Chandrasekhar2010} if we assume the specific heat $c_V$ (at constant volume) is a constant, for \textit{polytropic changes}, we have $\hat{\mathcal{T}}= \Theta_{\gamma^\prime}\hat{\rho}^{\gamma^\prime-1}$ and $p=R \mathcal{T} \rho$ (note $[R]=\frac{L^2}{T^2\mathbb{T}}$) where $\Theta_{\gamma^\prime}$ is the \textit{polytropic temperature}  and $\gamma^\prime=\frac{c_p-c}{c_V-c}$ is the polytropic exponent and $R=\bar{R}/\mathcal{M}$ is the specific gas constant and $\bar{R}=8.314$J/(mol$\cdot $K) is the universal gas constant and $\mathcal{M}$ is the molar mass. We then arrive at $p=R\mathcal{T}_T\Theta_{\gamma^\prime} \rho_T^{1-\gamma^\prime} \rho^{\gamma^\prime}$. In view of the expression of the entropy, 
\begin{equation}\label{e:Th1}
	S=S_0+c_V \bigl[\ln \Theta_{\gamma^\prime}+(\gamma^\prime-\gamma)\ln \hat{\rho}\bigr] \quad \Rightarrow  \quad  \Theta_{\gamma^\prime}=e^{\frac{S-S_0}{c_V}} \hat{\rho}^{\gamma-\gamma^\prime}, 
\end{equation}
where $\gamma=\frac{c_p}{c_V}=1+\frac{R}{c_V}$. 
We reexpress $p =K  e^{\frac{S-S_0}{c_V}}  \rho^{\gamma}  $ 
where $K=\bar{R} \mathcal{T}_T\rho_T^{1-\gamma} /\mathcal{M} $.  If $\gamma=4/3$, the nondimensionalized
equation of state becomes $\hat{p} = \hat{K} e^{s}  \hat{\rho}^{\frac{4}{3}}  $ where  denoting $s=(S-S_0)/c_V$, $\hat{K}=  \bar{R}\mathcal{T}_T /(\mathcal{M}  K_T \rho_T^{\frac{1}{3}})$.  By \eqref{e:Th1}, we have
\begin{equation*}\label{e:temeq}
	e^{s}=\hat{\mathcal{T}} \hat{\rho}^{-\frac{1}{3}}
\end{equation*}
Thus, the initial condition of entropy \eqref{e:datab} implies the initial distribution of the temperature $\mathcal{T} \propto |\boldsymbol{x}|^2$.

To simplify notation, we will drop the ``hat'' from the hatted variables throughout main body of this article.

\section{Mathematical preparation of a nonlinear ODE}\label{s:ODE} 
The main mathematical tool is a type of ODEs developed in our previous article \cite[\S$2$]{Liu2022b}. For readers' convenience, we quote the results without proofs in this appendix and readers can find detailed proof in \cite[\S$2$]{Liu2022b}. 
We consider the solutions $f(t)$ to the following type ODE, 
 \begin{gather}
 	f^{\prime\prime}(t)+\frac{\ca}{t}  f^\prime(t)-\frac{\cb}{t^2} f(t)(1+  f(t))-\frac{\cc (  f^\prime(t))^2}{1+f(t)}=   0 , \label{e:feq0}\\
 	f(t_0)= \mf>0 \AND
 f^\prime(t_0)=   \mf_0>0 ,  \label{e:feq1}
 \end{gather}
 where $\mf,\mf_0>0$ are positive constants and
 \begin{equation}\label{e:abcdk}
 	\ca>1, \quad \cb>0 \AND 1<\cc < 3/2 .   
 \end{equation}

From now on, to simplify the notations, we denote 
\begin{equation*}%\label{e:deftr}
	\triangle:=\sqrt{(1-\ca)^2+4\cb}>-\ba, \; \ba=1-\ca<0, \; \bc=1-\cc<0
\end{equation*}
and introduce constants $\mathtt{A}$,  $\mathtt{B}$, $\mathtt{C}$, $\mathtt{D}$ and $\mathtt{E}$ depending on the initial data  $\mf$ and $\mf_0$ to \eqref{e:feq0}--\eqref{e:feq1} and parameters $\ca$, $\cb$ and $\cc$,
\begin{align*}
	\mathtt{A}:=&\frac{t_0^{   -\frac{\ba-\triangle}{2}  }}{\triangle}\biggl(  \frac{t_0   \mf_0}{(1+\mf)^2} - \frac{\ba+\triangle}{2}  \frac{\mf  } {1+\mf} \biggr), \\
	\mathtt{B}:= &  \frac{t_0^{-\frac{\ba+\triangle}{2} }}{\triangle} \Bigl( \frac{\ba-\triangle}{2}  \frac{\mf }{1+\mf}  -\frac{t_0 \mf_0}{(1+\mf)^2} \Bigr)<0, \\
	\mathtt{C}:= & \frac{2 } {2+\ba+\triangle} \Bigl( \ln( 1+\mf) +\frac{\ba+\triangle}{2\cb} \frac{t_0\mf_0}{1+\mf}\Bigr)t_0^{-\frac{\ba+\triangle}{2 }} >0,  \\
	\mathtt{D}:= &
	\frac{ \ba+\triangle   }{2+\ba+\triangle}  \Bigl(  \ln( 1+\mf)  - \frac{1 } { \cb} \frac{t_0\mf_0}{1+\mf}\Bigr) t_0,   \\
	\mathtt{E}:=  &
	\frac{\bc  \mf_0 t_0^{1-\ba}  }{  \ba  (1+\mf) } >0.
\end{align*}

We define the following two critical times $t_\star$ and $t^\star$. 
	\begin{enumerate}
		\item
		Let $\mathcal{R}:=\{t_r>t_0 \;|\;\mathtt{A} t_r^{\frac{\ba-\triangle}{2} } + \mathtt{B} t_r^{\frac{\ba+\triangle}{2} } + 1 =0\}$ and define $t_\star:=\min \mathcal{R}$.
		\item If $t_0^{\ba}> \mathtt{E}^{-1}$, we define $t^\star :=    (t_0^{\ba}- \mathtt{E}^{-1} )^{1/\ba}\in(0,\infty)$, i.e.,  $t=t^\star$ solves $1-\mathtt{E}  t_0^{\ba} +  \mathtt{E}   t^{\ba}=0$.
	\end{enumerate}

We are now in a position to state the main theorem on ODE \eqref{e:feq0}--\eqref{e:feq1} and the proof can be found in \cite[\S$2$]{Liu2022b}. 
\begin{theorem}\label{t:mainthm0}
	Suppose constants $\ca$, $\cb$ and $\cc$ are defined by  \eqref{e:abcdk}, $t_\star$ and $t^\star$ are defined above and the initial data $\mf, \mf_0>0$, then
	\begin{enumerate}
		\item $t_\star \in[0,\infty)$ exists and $t_\star>t_0$. 
		\item There is a constant $t_m\in [t_\star,\infty]$, such that there is a unique solution $f\in C^2([t_0,t_m))$ to the equation \eqref{e:feq0}--\eqref{e:feq1}, and
		\begin{equation*}%\label{e:limf}
			\lim_{t\rightarrow t_m} f(t)=+\infty \AND \lim_{t\rightarrow t_m} f_0(t)=+\infty .
		\end{equation*}
		\item  $f$ satisfies upper and lower bound estimates,
		\begin{align*}%\label{e:fbds}
			1+f(t)>&\exp \bigl( \mathtt{C} t^{\frac{\ba+\triangle}{2} }  +\mathtt{D}  t^{-1}\bigr)      &&\text{for}\quad t\in(t_0,t_m);
			\\
			1+f(t) < & \bigl(\mathtt{A} t^{\frac{\ba-\triangle}{2} } + \mathtt{B} t^{\frac{\ba+\triangle}{2} } + 1 \bigr)^{-1}    && \text{for}\quad t\in(t_0,t_\star).
		\end{align*}
	\end{enumerate}		
	Furthermore, if the initial data satisfies  
	$\mf_0 >  \ba(1+\mf) /(\bc t_0 )$,
	then
	\begin{enumerate}
		\setcounter{enumi}{3}
		\item 	$t_\star$  and $t^\star$  exist and finite, and  $t_0<t_\star<t^\star<\infty$;
		\item There is a finite time $t_m\in [t_\star,t^\star)$, such that there is a solution $f\in C^2([t_0,t_m))$ to Eq.  \eqref{e:feq0} with the initial data \eqref{e:feq1},   and 	\begin{equation*}
			\lim_{t\rightarrow t_m} f(t)=+\infty \AND \lim_{t\rightarrow t_m} f_0(t)=+\infty .
		\end{equation*}
		\item The solution $f$ has improved lower bound estimates, for $t\in(t_0,t_m)$,
		\begin{equation*}\label{e:ipvest}
			(1+\mf)  \bigl(1-\mathtt{E}  t_0^{\ba} +  \mathtt{E}   t^{\ba} \bigr)^{1/\bc}  < 1+f(t) .
		\end{equation*}
	\end{enumerate}
\end{theorem}

%-----------------------------------------------------------------------New Subsec-----------------------------------------------------------------------

\section*{Acknowledgement}
		%We would like to thank ... for their helpful discussions, comments and advice.
		C.L. is supported by the Fundamental Research Funds for the Central Universities, HUST: $5003011036$, $5003011047$, and the National Natural Science Foundation of China (NSFC) under
		the Grant No. $11971503$.
		%We also thank the referee for their comments and criticisms, which have served to improve the content and exposition of this article.

\bibliographystyle{unsrt}
\bibliography{Reference_Chao}

\begin{thebibliography}{10}

\bibitem{Liu2022b}
Chao Liu.
\newblock Blowups for a class of second order nonlinear hyperbolic equations: A
  reduced model of nonlinear jeans instability.
\newblock {\em arXiv:2208.06788}, August 2022.

\bibitem{Liu2023b}
Chao Liu.
\newblock Fully nonlinear gravitational instabilities for expanding spherical
  symmetric newtonian universes with inhomogeneous density and pressure.
\newblock {\em arXiv:2305.13211}, 2023.

\bibitem{Jeans1902}
J.~H. Jeans.
\newblock The stability of a spherical nebula.
\newblock {\em Philos. Trans. R. Soc. Lond. A}, 199:1--53, 1902.

\bibitem{Lifshitz1946}
E.~M. {Lifshitz}.
\newblock {On the gravitational stability of the expanding universe}.
\newblock {\em Zhurnal Eksperimentalnoi i Teoreticheskoi Fiziki}, 16:587--602,
  January 1946.

\bibitem{Bonnor1957}
W.~B. Bonnor.
\newblock Jeans' formula for gravitational instability.
\newblock {\em Monthly Notices of the Royal Astronomical Society},
  117(1):104--117, feb 1957.

\bibitem{Zeldovich1971}
Ya.~B. Zel'dovich and I.~D. Novikov.
\newblock {\em Relativistic astrophysics 2: The Structure and Evolution of the
  Universe}.
\newblock University of Chicago Press, Chicago, 1971.

\bibitem{ViatcehslavMukhanov2013}
Viatcheslav Mukhanov.
\newblock {\em Physical foundations of cosmology}.
\newblock Cambridge University Press, November 2013.

\bibitem{Liu2022}
Chao Liu and Yiqing Shi.
\newblock Rigorous proof of the slightly nonlinear {Jeans} instability in the
  expanding {Newtonian} universe.
\newblock {\em Physical Review D}, 105(4):043519, feb 2022.

\bibitem{Rendall2002}
Alan~D. Rendall.
\newblock Theorems on existence and global dynamics for the {Einstein}
  equations.
\newblock {\em Living Reviews in Relativity}, 5(1), sep 2002.

\bibitem{Peebles2020}
P.~J.~E. Peebles.
\newblock {\em The Large-Scale Structure of the Universe}.
\newblock Princeton University Press, September 2020.

\bibitem{Arbuzova2014}
E.V. Arbuzova, A.D. Dolgov, and L.~Reverberi.
\newblock Jeans instability in classical and modified gravity.
\newblock {\em Physics Letters B}, 739:279--284, dec 2014.

\bibitem{Sciama1955}
D.~W. Sciama.
\newblock On the formation of galaxies in a steady state universe.
\newblock {\em Monthly Notices of the Royal Astronomical Society},
  115(1):3--14, feb 1955.

\bibitem{LANDAU1975}
L.~D. Landau and E.~M. Lifshitz.
\newblock {\em The classical theory of fields}.
\newblock Oxford: Pergamon Press, 1975, 4th rev.engl.ed.

\bibitem{DeGroot2012}
P.~Mazur S.~R. De~Groot.
\newblock {\em Non-Equilibrium Thermodynamics}.
\newblock Dover Publications, December 2012.

\bibitem{Wang2005}
Zhuxi Wang.
\newblock {\em Thermodynamics}.
\newblock Peking University Press, 2nd edition, 2005.
\newblock In Chinese.

\bibitem{Fosalba1998}
Pablo Fosalba and Enrique Gazta{\~{n}}aga.
\newblock Cosmological perturbation theory and the spherical collapse model
  {\textemdash} i. gaussian initial conditions.
\newblock {\em Monthly Notices of the Royal Astronomical Society},
  301(2):503--523, dec 1998.

\bibitem{Frauendiener1995}
J.~Frauendiener and C.~Klein.
\newblock On crossing dust shells.
\newblock {\em Journal of Mathematical Physics}, 36(7):3632--3643, jul 1995.

\bibitem{Szekeres1999}
Peter Szekeres and Anthony Lun.
\newblock What is a shell-crossing singularity?
\newblock {\em The Journal of the Australian Mathematical Society. Series B.
  Applied Mathematics}, 41(2):167--179, oct 1999.

\bibitem{Chandrasekhar2010}
Subrahmanyan Chandrasekhar.
\newblock {\em An Introduction to the Study of Stellar Structure}.
\newblock Dover Publications, 2010.

\end{thebibliography}
	
\end{document}